\documentclass{article}
\usepackage{spconf,amsmath,graphicx}
\usepackage{booktabs,adjustbox,makecell,multirow}

\usepackage[labelformat=simple]{subcaption}

\usepackage[font=small,labelfont=bf]{caption}
\usepackage{flushend}

\usepackage{pifont}
\newcommand{\cmark}{\ding{51}}%
\newcommand{\xmark}{\ding{55}}%

\title{Integration of speech separation, diarization, and recognition for multi-speaker meetings:
System description, Comparison, and Analysis
}
\name{\begin{tabular}{c}Desh Raj$^1$, Pavel Denisov$^2$, Zhuo Chen$^3$, Hakan Erdogan$^4$, Zili Huang$^1$, Maokui He$^5$, Shinji Watanabe$^1$,\\ Jun Du$^5$, Takuya Yoshioka$^3$, Yi Luo$^6$, Naoyuki Kanda$^3$, Jinyu Li$^3$, Scott Wisdom$^4$, John R. Hershey$^4$\end{tabular}}
\address{$^1$Center for Language and Speech Processing, The Johns Hopkins University, Baltimore, MD\\$^2$Institute for Natural Language Processing, University of Stuttgart, Germany\\$^3$Microsoft Corp, Redmond, WA, $^4$Google Research, Cambridge, MA\\$^5$University of Science and Technology of China, HeFei, China\\Department of Electrical Engineering, Columbia University, NY}
\email{draj@cs.jhu.edu}

\vspace{-1em}
\begin{document}
%
\maketitle
\begin{abstract}

Multi-speaker speech recognition of unsegmented recordings has diverse applications such as meeting transcription and automatic subtitle generation. With technical advances in systems dealing with speech separation, speaker diarization, and automatic speech recognition (ASR) in the last decade, it has become possible to build pipelines that achieve reasonable error rates on this task. In this paper, we propose an end-to-end modular system for the LibriCSS meeting data, which combines independently trained separation, diarization, and recognition components, in that order. We study the effect of different state-of-the-art methods at each stage of the pipeline, and report results using task-specific metrics like SDR and DER, as well as downstream WER. Experiments indicate that the problem of overlapping speech for diarization and ASR can be effectively mitigated with the presence of a well-trained separation module. Our best system achieves a speaker-attributed WER of 12.7\%, which is close to that of a non-overlapping ASR. 

\end{abstract}
\section{Introduction}
\label{sec:intro}

\begin{figure*}
\centering
\includegraphics[width=0.8\linewidth]{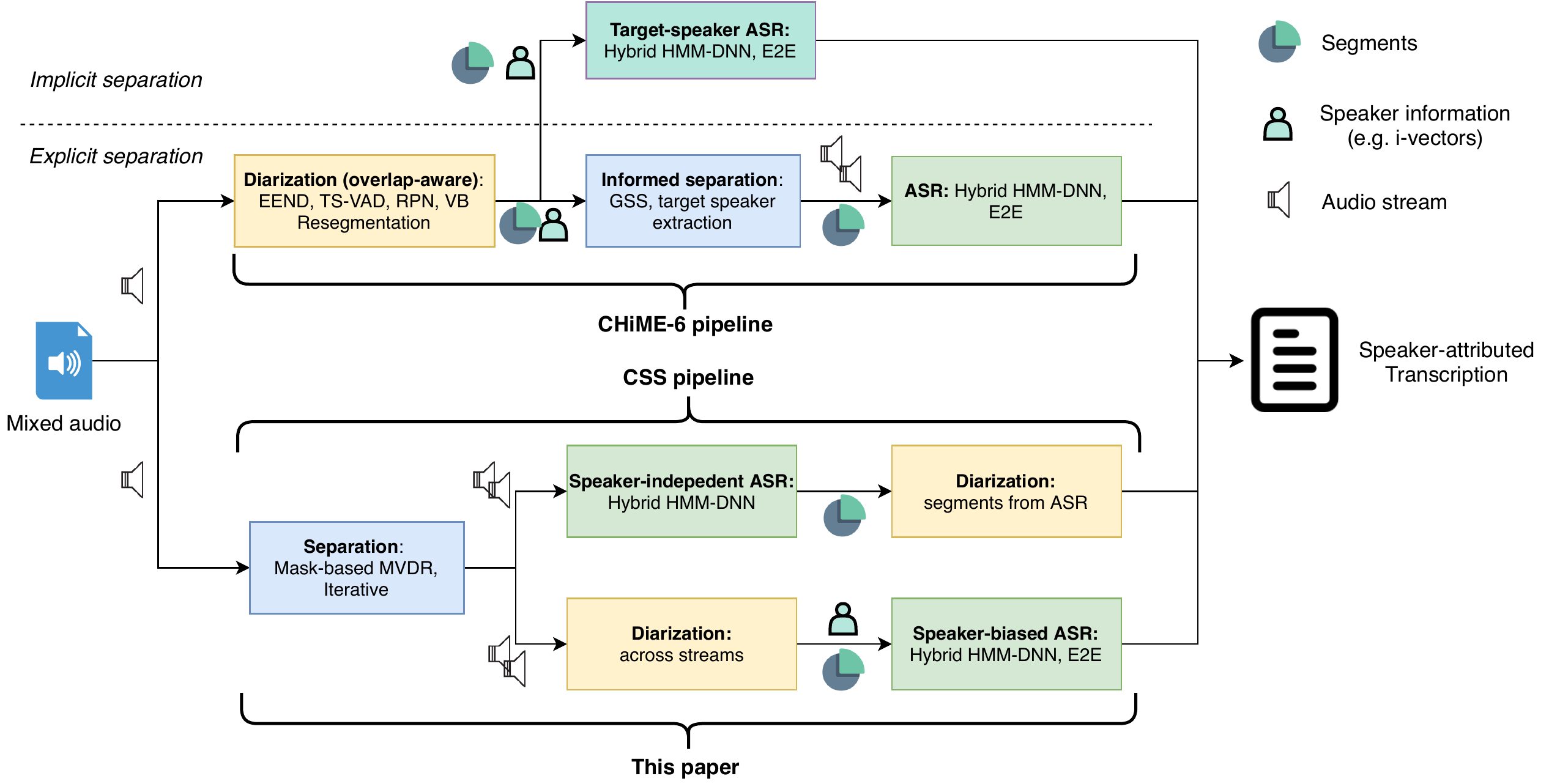}
\caption{An overview of our proposed approach, compared with the CHiME-6 pipeline~\cite{Watanabe2020CHiME6CT} and the CSS pipeline~\cite{Chen2020ContinuousSS}.}
\label{fig:overview}
\vspace{-1em}
\end{figure*}

Multi-speaker speech recognition is defined as a task wherein, given a long unsegmented recording consisting of a conversation between an unknown number of participants, the expected output contains ``who said what and when.'' 
Although the last two decades have seen incredible leaps in speech technology --- deep learning systems have matched human parity in the single-speaker conversational speech settings~\cite{Xiong2016AchievingHP} --- multi-speaker speech processing remains a challenge as such recordings may contain almost 20\% overlapped speech~\cite{Carletta2005TheAM,Shriberg2001ObservationsOO}. This scenario has implications for both diarization and automatic speech recognition (ASR) --- single-speaker diarization systems miss the interfering speaker completely, and ASR systems trained on clean utterances are more error-prone on overlapped regions. These factors, along with the effect of far-field acoustic conditions, may increase the speaker-attributed word error rates by up to 86\%~\cite{Yoshioka2019MeetingTU} in meetings. To tackle such problems, recent editions of DIHARD~\cite{Ryant2019TheSD} and CHiME~\cite{Watanabe2020CHiME6CT} have focused on these tasks in very challenging settings.

A straightforward approach for solving the overlap problem in multi-speaker conversations is through speech separation. Deep learning based separation methods have been progressing consistently over the last few years --- a signal-to-distortion ratio of 19.0 dB has been obtained on the popular WSJ0-2mix dataset~\cite{Luo2020DualPathRE}, making the enhanced signals almost indistinguishable from clean utterances~\cite{Luo2019ConvTasNetSI}. However, these techniques are often evaluated on short, fully overlapping (and often simulated) mixtures, and do not measure distortions introduced in overlap-free regions. They may also make unrealistic assumptions, such as prior knowledge of the number of speakers in the mixture. Recently, there have been efforts towards ``continuous speech separation,'' which situates separation techniques in more realistic settings of long-form conversations containing partially overlapped speech~\cite{Yoshioka2019LowlatencySC,Chen2020ContinuousSS}, such as those found in multi-speaker conversations.

There is increasing interest towards combining the advances made in speech separation, diarization, and ASR for tackling multi-speaker speech recognition. The CHiME-6 challenge~\cite{Watanabe2020CHiME6CT} included a track aimed at recognizing unsegmented dinner-party conversations recorded on multiple microphone arrays, and baseline evaluations indicated that replacing oracle segments with diarization outputs could degrade downstream WERs by almost 52\%. Chen et al.~\cite{Chen2020ContinuousSS} proposed the new LibriCSS meeting dataset, and combined speaker-independent continuous speech separation~\cite{Yoshioka2019LowlatencySC} with a strong hybrid ASR~\cite{Lu2019PyKaldi2YA} to recognize long unsegmented meeting recordings. Similar datasets such as LibriMix~\cite{Cosentino2020LibriMixAO} have been created to satisfy the need for multi-speaker conversational data containing partial overlaps. Joint modeling of speaker counting, speaker identification, and ASR using end-to-end models has also been investigated~\cite{Kanda2020JointSC,k2020investigation}.

Nevertheless, the number of available methods to perform separation, diarization, or recognition are plenty --- each with their own sets of requirements and assumptions. Together with the several possibilities of combining them, this creates a daunting task to analyze systems that recognize unsegmented multi-speaker recordings. In this paper, we make a first attempt at tackling this problem, using the LibriCSS dataset. We propose a novel end-to-end modular pipeline that performs separation, diarization, and recognition, in that order. For each of these modules, we compare several existing methods, describing the advantages and disadvantages of each alternative. We also demonstrate the impact of separation on downstream tasks by presenting corresponding results on the original mixed recording. Since the modules only interact at the I/O level, they can be implemented using different tools and optimized independently. Our best system with and without separation achieves a concatenated minimum-permutation WER (cpWER)~\cite{Watanabe2020CHiME6CT} of 12.7\% and 23.9\%, respectively, on the LibriCSS evaluation set. We will release code to reproduce (and extend) our pipeline here: {\footnotesize \texttt{https://desh2608.github.io/pages/jsalt/}}.

\vspace{-0.6em}
\section{System Overview}
\label{sec:overview}

Our pipeline consists of three components: Separation $\rightarrow$ Diarization $\rightarrow$ ASR. An unsegmented multichannel audio recording is provided as input to a continuous speech separation system, such as~\cite{Yoshioka2019LowlatencySC}. The separation module works on small windowed segments containing at most 2 or 3 speakers, and produces the corresponding number of separated audio streams, which are passed to the diarization module. Since a speaker may have been split into different streams across different windows, the diarization is performed by considering all the audio streams simultaneously. We will discuss the implications of this requirement on different diarization methods in Section~\ref{sec:diarization}. After diarization, the single-speaker homogenenous segments are fed into an ASR decoder. 

Fig.~\ref{fig:overview} shows our proposed approach, and situates it in comparison with alternative approaches proposed in \cite{Watanabe2020CHiME6CT} and \cite{Chen2020ContinuousSS}, which we refer to as the \textit{CHiME-6 pipeline} and the \textit{CSS pipeline}, respectively. All 3 methods perform explicit separation, as opposed to an implicit ``separation+recognition'' usually performed by target-speaker ASR methods~\cite{Delcroix2018SingleCT}. Stacking the components in different orders elicits unique advantages and limitations for each approach. Since diarization is performed first in the CHiME-6 pipeline, it makes it possible to use the source activity pattern derived from the diarization output to guide the estimation of the mixture model parameters in guided source separation (GSS)~\cite{Bddeker2018FrontendPF}, or use the speaker information to perform target speaker extraction, as in SpeakerBeam~\cite{molkov2019SpeakerBeamSA}. However, poor diarization can have significant impacts on separation results in this pipeline. The CSS pipeline performs separation in the first stage, followed by ASR and diarization, allowing the system to be deployed in a streaming setting. The diarization performance may itself also benefit from reduced false alarms, since segments are obtained from the ASR output. However, it makes speaker-biasing infeasible at the ASR stage. Furthermore, the ASR output may contain spurious insertions resulting from cross-talk in non-overlap regions of the separated audio streams.

In contrast, our pipeline makes it feasible to employ a speaker-biased ASR, and also to remove extra insertions through simple post-processing of the diarization output (cf. Section~\ref{sec:sep_results}). Unlike the CHiME-6 pipeline, our separation is not dependent on the performance of the diarizer, and speaker-independent separation techniques can be effectively used in the pipeline. Here we demonstrate two of these --- mask-based minimum variance distortionless response (MVDR), and sequential multi-frame separation. Furthermore, since the diarization stage sees only single-speaker recordings, it circumvents the need for an overlap-aware diarization component. 

We will now describe the performance of different methods in each module -- namely separation, diarization, and ASR --- in Sections~\ref{sec:separation}, \ref{sec:diarization}, and \ref{sec:asr}, respectively. All our experiments are conducted on LibriCSS~\cite{Chen2020ContinuousSS}, which consists of multi-channel audio recordings of ``simulated conversations,'' generated by mixing Librispeech utterances~\cite{Panayotov2015LibrispeechAA}. It comprises 10, approximately one-hour long, sessions. Each session is made up of six 10-minute-long ``mini sessions'' that have different overlap ratios, ranging from 0\% to 40\%. The recordings were made in a regular meeting room by using a seven-channel circular microphone array. We used \textit{session 0} as the development set, and the remaining 9 sessions for evaluation (which we use to report results). Diarization results are reported using diarization error rate (DER), and ASR performance is evaluated in terms of cpWER. It is computed by concatenating all the utterances of a speaker in the reference and hypothesis, scoring all speaker pairs, and then finding the speaker permutation that minimizes the total WER.

\vspace{-0.6em}
\section{Speech Separation}
\label{sec:separation}

\subsection{Mask-based MVDR}
\label{sec:mvdr}

The speech separation module follows the continuous speech separation scheme~\cite{Yoshioka2018MultiMicrophoneNS}, consisting of three steps. First, the recording is uniformly segmented into smaller chunks with overlap. For each chunk, a multi-channel separation network~\cite{Chen2020ContinuousSS}, trained with permutation invariant training (PIT) criterion, is applied to estimate three time-frequency (TF) masks: two for speech sources and one for noise. In this scheme, we used a chunk size of 2.4s with 0.8s hop.

Thereafter, a stitching algorithm~\cite{kolbaek2017multitalker} is used to track local permutations and glue the masks from each chunk into the meeting-wise mask. This stitching is performed using mask similarity between adjacent chunks on the overlapped region (which is 1.6s in our case), by finding the permutation that has minimum distance between chunks. 
Based on the resulting permutation, the chunk-wise masks are connected to obtain a mask stream for the long recording. We averaged the overlap region between chunks for this process.

Finally, given the stitched masks for the entire meeting, a mask-based adaptive MVDR beamforming~\cite{Yoshioka2019AdvancesIO} is performed to get the final separation result. Since the noise is mostly stationary, we used the noise mask for the entire meeting to estimate the noise spatial covariance. 

Note that the number of output channels in each local separation is highly correlated with the chunk size. As \cite{Yoshioka2019AdvancesIO} suggests, most 2.4s chunks contains at most 2 speakers, therefore two output channels suffice for local processing. However, when the chunk size is larger (as in the next separation model), it may contain more than 2 speakers, even though LibriCSS contains at most 2-speaker overlaps. 

\vspace{-0.5em}
\subsection{Sequential multi-frame separation}

We also used a novel sequential separation system  which has multiple mask-based beamforming steps \cite{wang2021sequential}. A mask-based multi-frame multi-channel Wiener filter (MCWF) beamformer is used. We used three untied sequential steps inside the model by feeding previously beamformed signals into the next mask-prediction step. The output of this model can be the last mask-network output or the one previous beamformed output. This model was trained using 10 second long mixtures of three speaker utterances from Libri-Light database~\cite{Kahn2020LibriLightAB}. During training, the utterances were mixed on-the-fly with room impulse responses obtained using an image-method based room simulator that simulates data acquisition in shoe-box shaped rooms with an 8-microphone array \cite{wang2021sequential}. This pre-trained model was used to separate sources in the LibriCSS dev and eval datasets using 8 second long blocks with 4 seconds overlap. Since LibriCSS has 7 microphones, we added another microphone signal by shifting the first microphone signal with one sample and adding white Gaussian noise with variance 1e-6. The source estimate outputs were stitched using a stitching loss of magnitude STFT domain mean-squared error between time-domain signals to resolve the permutation across blocks. This is similar to the stitching method described in Section~\ref{sec:mvdr}.

\vspace{-0.5em}
\subsection{Experimental results}
\label{sec:sep_results}

Table~\ref{tab:sep_result} shows the performance of the separation methods on the LibriCSS eval set. Separation performance is calculated on a simulated eval set different from LibriCSS since reference signals are required. Since the models work with varying window sizes, we first mapped the output tracks to the $N$ source tracks for every 0.8s block\footnote{0.8s is the highest common factor between the chunk size of the models.}, where $N$ is the number of participating speakers, using magnitude-domain distance and assuming at most two active speakers per 0.8s block, and then generated $N$ speaker tracks from the separated outputs. We call this process ``oracle track mapping''. We report separation performance in terms of average meeting-level SDR~\cite{vincent2006performance}. Sequential multi-frame model achieved a higher SDR of 14.1 dB as compared to the mask-based MVDR's 5.8 dB, since implemented MVDR beamformer does not attempt to reconstruct target signals exactly.

We also evaluated the separated audio with a spectral clustering based diarizer (Section~\ref{sec:spectral}) and a hybrid HMM-DNN ASR (Section~\ref{sec:hybrid}), and the results are shown in the table in terms of DER and cpWER, respectively. 
From the table, we note that the diarization performance of the two methods were comparable, but the 3-stream sequential multi-frame separation outperformed the 2-stream mask-based MVDR in terms of cpWER results. Although the sequential model was trained with an 8-microphone cubic geometry, it generalized well to LibriCSS, which has a completely different geometry, since the model is inherently geometry-independent. 

Furthermore, a simple post-processing trick applied on the diarization output was found to be useful for the sequential model -- using this trick improved the cpWER by 15.4\% relative. For this post-processing, we removed a segment from a stream if it was completely enclosed within a same-speaker segment in a different stream. This is akin to filtering out cross-talk prior to ASR decoding. This trick was not useful for the mask-based MVDR, providing only a marginal improvement of 0.2\% relative in terms of cpWER. Note that although sequential model achieved significantly better SDR, their cpWER are similar. This indicates the potential objective mismatch between speech separation and recognition.

\begin{table}[t]
\centering
\caption{Performance of separation methods on LibriCSS eval set in terms of resulting downstream diarization (using spectral clustering) and ASR (using TDNN-F model) results. For comparison, we also show results obtained on a ``no separation'' baseline. Separation performance is reported on a simulated eval set. $^{\dag}$MVDR beamformer does not attempt to reconstruct the target signal at the reference microphone directly, so the separation metric reflects this mismatch.}
\label{tab:sep_result}
\begin{adjustbox}{max width=\linewidth}
\begin{tabular}{@{}lccc@{}}
\toprule
\textbf{Method} & \textbf{SDR (dB)} & \textbf{DER (\%)} & \textbf{cpWER (\%)} \\ \midrule
No separation & - & 18.3 & 31.0 \\
Mask-based MVDR & 5.8$^{\dag}$ & \textbf{13.9} & 22.8 \\
Sequential multi-frame & \textbf{14.1} & 14.1 & \textbf{19.3} \\ \bottomrule
\end{tabular}
\end{adjustbox}
\end{table}

\section{Speaker Diarization}
\label{sec:diarization}
\vspace{-0.2em}

We can categorize diarization methods based on whether or not they can assign overlapping speaker segments. 
Since our pipeline separates the recording prior to diarization, overlap-awareness is not a strict requirement. At the same time, since the separation is done on windowed segments, the diarization needs to be performed across audio streams (since a speaker can be present in different streams at different times). Based on these conditions, we selected the following diarization methods for our study. 

\vspace{-0.5em}
\subsection{X-vector + clustering}
\label{sec:spectral}
\vspace{-0.2em}

This method consists of a speech activity detection (SAD) component followed by clustering of small subsegment embeddings. We used a similar SAD as that described in \cite{Watanabe2020CHiME6CT}, consisting of a TDNN-Stats based classifier with Viterbi decoding for inference. The speech segments were divided into subsegments with a window size of 1.5s and a stride of 0.75s, and 128-dimensional embeddings were extracted using an x-vector extractor~\cite{Snyder2018XVectorsRD} trained on VoxCeleb data~\cite{Nagrani2017VoxCelebAL} with simulated room impulse response~\cite{Ko2017ASO}. We conducted experiments with 3 variants of clustering: (i) agglomerative hierarchical clustering (AHC) on PLDA scores~\cite{GarciaRomero2017SpeakerDU}, (ii) spectral clustering (SC) on cosine similarity~\cite{Park2020AutoTuningSC}, and (iii) VBx clustering initialized from the AHC system~\cite{Diez2019BayesianHB}. For these methods, we did not make any assumptions about the number of speakers in the recording. We used the same PLDA trained on Librispeech for both AHC and VBx, and did not use the PLDA interpolation technique from \cite{Diez2019BayesianHB}.

\vspace{-0.5em}
\subsection{Region proposal networks (RPN)}
\vspace{-0.2em}

This is a supervised method, which combines the segmentation and embedding extraction steps into a single neural network and jointly optimizes them~\cite{Huang2020SpeakerDW}. The region embeddings are then clustered (using K-means clustering) and a non-maximal suppression is applied. We trained the RPN on simulated meeting-style recordings with partial overlaps generated using utterances from the Librispeech~\cite{Panayotov2015LibrispeechAA} training set. Since we used K-means clustering, we assumed that the oracle number of speakers for each recording is known.

\vspace{-0.5em}
\subsection{Target-speaker voice activity detection (TS-VAD)}
\vspace{-0.2em}

The TS-VAD model takes conventional speech features (e.g., MFCC) along with i-vectors for each speaker as inputs and produces frame-level activities for each speaker using a neural network with a set of binary classification output layers~\cite{Medennikov2020TargetSpeakerVA}. Since the number of these binary output nodes are fixed for training, we assumed that the maximum possible number of speakers in any session is at most 8. The initial estimates for the speaker i-vectors were obtained using the SC system. For training, we created simulated meeting-style data similar to that used for training the RPN model.

\vspace{-0.5em}
\subsection{Experimental results}
\vspace{-0.2em}

We conducted experiments with ``mixed'' as well as ``separated'' audio to analyze the impact of separation on diarization performance. For the mixed recording, we selected the first channel as our input. For evaluation on separated streams, we fixed the separation component as mask-based MVDR, and the ASR was chosen to be a speaker-biased hybrid HMM-DNN (described in Section~\ref{sec:hybrid}). Table~\ref{tab:mixed_diar} shows the diarization performance on mixed LibriCSS, with a breakdown by overlap condition. The SAD error for the clustering-based systems was 4.8\%. It is immediately evident that assigning overlapping speech is important to perform well on this task -- both RPN and TS-VAD outperformed clustering-based methods. Even on low overlap regions, there is a significant difference, and we conjecture that this arises from a mismatch between the training data for the PLDA used for scoring the AHC and VBx models, and the evaluation set. Since SC uses cosine scoring, it performed better than the other clustering methods. For the RPN and TS-VAD systems, creating a simulated mixture which closely resembled the test set was found to be important, and using cepstral mean normalization (CMN) was critical for this performance. 

\begin{table}[t]
\centering
\caption{Diarization performance on mixed LibriCSS evaluation set, in terms of \% DER. 0S and 0L refer to 0\% overlap with short and long inter-utterance silences, respectively. RPN and TS-VAD methods assume that oracle number of speakers is known. We also report the corresponding cpWER using a TDNN-F based ASR model. Using an oracle diarization output results in a cpWER of 23.1\%.}
\label{tab:mixed_diar}
\begin{adjustbox}{max width=\linewidth}
\begin{tabular}{@{}lcccccccc@{}}
\toprule
\multicolumn{1}{l}{\multirow{2}{*}{\textbf{Method}}} & \multicolumn{6}{c}{\textbf{Overlap ratio in \%}} & \multicolumn{1}{c}{\multirow{2}{*}{\textbf{DER}}} & \multicolumn{1}{c}{\multirow{2}{*}{\textbf{cpWER}}} \\
\cmidrule(r{4pt}){2-7}
\multicolumn{1}{c}{} & \multicolumn{1}{c}{\textbf{0L}} & \multicolumn{1}{c}{\textbf{0S}} & \multicolumn{1}{c}{\textbf{10}} & \multicolumn{1}{c}{\textbf{20}} & \multicolumn{1}{c}{\textbf{30}} & \multicolumn{1}{c}{\textbf{40}} & \multicolumn{1}{c}{} \\
\midrule
AHC & 16.1 & 12.0 & 16.9 & 23.6 & 28.3 & 33.2 & 22.6 & 36.7 \\
VBx & 14.6 & 11.1 & 14.3 & 21.5 & 25.4 & 31.2 & 20.5 & 33.4 \\
SC & 10.9 & 9.5 & 13.9 & 18.9 & 23.7 & 27.4 & 18.3 & 31.0 \\
RPN & \textbf{4.5} & 9.1 & 8.3 & \textbf{6.7} & 11.6 & 14.2 & 9.5 & 27.2 \\
TS-VAD & 6.0 & \textbf{4.6} & \textbf{6.6} & 7.3 & \textbf{10.3} & \textbf{9.5} & \textbf{7.6} & \textbf{24.4} \\
\bottomrule
\end{tabular}
\end{adjustbox}
\end{table}

\begin{table}[t]
\centering
\caption{Comparison of diarization and downstream ASR performance of different methods on separated audio streams for LibriCSS evaluation set. We used mask-based MVDR for separation and TDNN-F based ASR for these experiments.}
\label{tab:sep_diar}
\begin{adjustbox}{max width=\linewidth}
\begin{tabular}{@{}clccccc@{}}
\toprule
\textbf{Separation} & \textbf{Metric} & \multicolumn{1}{c}{AHC} & \multicolumn{1}{c}{VBx} & \multicolumn{1}{c}{SC} & \multicolumn{1}{c}{RPN} & \multicolumn{1}{c}{TS-VAD} \\ \midrule
\xmark & \textbf{DER} & 22.6 & 20.5 & 18.3 & 9.5 & \textbf{7.6} \\
\xmark & \textbf{cpWER} & 36.7 & 33.4 & 31.0 & 27.2 & \textbf{24.4} \\ \midrule
\cmark & \textbf{DER} & 37.5 & - & \textbf{13.9} & 22.4 & 31.7 \\
\cmark & \textbf{cpWER} & 86.7 & - & \textbf{22.8} & 34.9 & 47.7 \\ \bottomrule
\end{tabular}
\end{adjustbox}
\end{table}

Next, we evaluated the systems on separated audio streams obtained using the mask-based MVDR method. Table~\ref{tab:sep_diar} shows these results, along with the downstream cpWER obtained using a hybrid HMM-DNN ASR model. Since AHC and SC perform subsegment-level clustering, they can be naturally extended to diarization across streams. VBx, on the other hand, estimates speaker changes through HMM state transitions, so it is not directly applicable to this scenario. RPN and TS-VAD can also be extended to this new setting, since RPN uses clustering of region embeddings, and TS-VAD predicts frame-level speaker activity based on the corresponding i-vectors. Our first observation is that among clustering-based methods, SC performed significantly better than AHC, likely because the PLDA used for AHC was trained on clean Librispeech utterances, which are acoustically very different from the separated audio. To verify this, we performed the AHC also on a cosine similarity matrix, and it resulted in an absolute DER improvement of 13.8\%. 
For RPN and TS-VAD, performance on mixed recording was not indicative of results on separated audio. Without any post-processing, we obtained a DER of 26.9\% using RPN. This improved to 22.4\% on filtering out non-speech segments using our SAD from Section~\ref{sec:spectral}. Similarly, TS-VAD performance degraded severely on going from mixed to separated audio. We found this degradation to be consistent for missed speech, false alarms, and speaker confusions, indicating a likely mismatch in train vs. test conditions. Consequently, the cpWER for both RPN and TS-VAD was found to be significantly higher than that for SC.

\section{Speech Recognition}
\label{sec:asr}

We conducted our ASR experiments on a hybrid TDNN-F based, and a Transformer-based end-to-end ASR model. Pretrained models (trained on Librispeech) for both of these are publicly available, enabling reproducibility of our results.

\begin{table}[t]
\centering
\caption{ASR performance on mixed LibriCSS evaluation set, in terms of \% WER. These results correspond to the ``utterance-wise evaluation'' of \cite{Chen2020ContinuousSS}, since we used oracle segments. The TDNN-F and Transformer E2E models obtain WERs of 8.8\% and 5.5\% (beam size 60), respectively, on the Librispeech \texttt{test-other} evaluation set.}
\label{tab:mixed_asr}
\begin{adjustbox}{max width=\linewidth}
\begin{tabular}{@{}lccccccc@{}}
\toprule
\multicolumn{1}{c}{\multirow{2}{*}{\textbf{Model}}} & \multicolumn{6}{c}{\textbf{Overlap ratio in \%}} & \multicolumn{1}{c}{\multirow{2}{*}{\textbf{Average}}} \\
\cmidrule(r{4pt}){2-7}
\multicolumn{1}{c}{} & \multicolumn{1}{c}{\textbf{0L}} & \multicolumn{1}{c}{\textbf{0S}} & \multicolumn{1}{c}{\textbf{10}} & \multicolumn{1}{c}{\textbf{20}} & \multicolumn{1}{c}{\textbf{30}} & \multicolumn{1}{c}{\textbf{40}} & \multicolumn{1}{c}{} \\
\midrule
TDNN-F (\textit{base}) & 16.1 & 16.0 & 27.7 & 39.1 & 49.4 & 58.3 & 36.8 \\
 + \textit{fine-tuned} & 11.4 & 11.4 & 18.2 & 26.3 & 33.7 & 40.7 & 25.2 \\
Transformer E2E (beam = 5) & 5.5 & 5.6 & 13.1 & 21.6 & 31.1 & 41.6 & 21.6 \\
Transformer E2E (beam = 30) & \textbf{5.1} & \textbf{5.3} & \textbf{11.4} & \textbf{19.5} & \textbf{28.5} & \textbf{38.3} & \textbf{19.7} \\
\bottomrule
\end{tabular}
\end{adjustbox}
\end{table}

\begin{table*}[t]
\centering
\caption{ASR performance on separated LibriCSS evaluation set, in terms of \% cpWER. These results correspond to the ``continuous-input evaluation'' of \cite{Chen2020ContinuousSS}. Mask-based MVDR and spectral clustering were used for separation and diarization, respectively.}
\label{tab:asr_sep}
\begin{adjustbox}{max width=\linewidth}
\begin{tabular}{@{}lcccccccccccccc@{}}
\toprule
\multirow{2}{*}{\textbf{Model}} & \multicolumn{7}{c}{\textbf{Without separation} (DER = 18.3\%)} & \multicolumn{7}{c}{\textbf{With separation} (DER = 13.9\%)} \\
\cmidrule(r{4pt}){2-8}\cmidrule(l){9-15}
 & \textbf{0L} & \textbf{0S} & \textbf{OV10} & \textbf{OV20} & \textbf{OV30} & \textbf{OV40} & \textbf{Average} & \textbf{0L} & \textbf{0S} & \textbf{OV10} & \textbf{OV20} & \textbf{OV30} & \textbf{OV40} & \textbf{Average} \\
\midrule
\textbf{TDNN-F} & 17.6 & 19.1 & 24.4 & 33.0 & 38.4 & 45.2 & 31.0 & 16.9 & 15.5 & 18.8 & 22.7 & 26.6 & 29.4 & 22.3 \\
\textbf{Transformer} & 12.4 & 14.1 & 20.2 & 29.5 & 35.3 & 41.9 & 27.1 & 11.2 & 8.5 & 10.6 & 14.8 & 15.5 & 17.5 & 13.4 \\
\bottomrule
\end{tabular}
\end{adjustbox}
\vspace{-1em}
\end{table*}

\vspace{-0.4em}
\subsection{TDNN-F based hybrid HMM-DNN}
\label{sec:hybrid}

Following the Kaldi~\cite{Povey2011TheKS} Librispeech recipe, we trained a 17-layer deep neural network consisting of factored TDNN layers~\cite{Povey2018SemiOrthogonalLM} using the lattice-free MMI objective~\cite{Povey2016PurelySN}. We used 40-dim MFCC features, and additionally appended 100-dim i-vectors estimated online. The model was trained on the 960h Librispeech data with 3x speed perturbation. We call this our \textit{base} model. On the Librispeech \texttt{test-clean} and \texttt{test-other} evaluation sets, this model obtains WERs of 3.8\% and 8.8\%, respectively. We additionally fine-tuned it for 1 epoch on Librispeech train set augmented with simulated room impulse responses~\cite{Ko2017ASO} to match the acoustic conditions of mixed LibriCSS recordings; this is referred to as the \textit{fine-tuned} model. For decoding on LibriCSS, we used the 3-gram language model provided with the Librispeech data, and the lattices were rescored using a pruned TDNN-LSTM based RNNLM trained on the training transcripts~\cite{Xu2018APR}. We used a 2-pass decoding strategy, where the i-vectors were re-estimated from the non-silence regions for the second pass~\cite{Manohar2019AcousticMF}. The fine-tuned model was used to evaluate the downstream ASR performance of the separation and diarization methods. 

\vspace{-0.4em}
\subsection{Transformer-based end-to-end ASR}
\label{sec:e2e}

Our end-to-end (E2E) ASR is a state-of-the-art ESPNet-based~\cite{Watanabe2018ESPnetES} Transformer encoder-decoder model~\cite{karita2019comparative}. It was trained on the 960h Librispeech corpus with SpecAugment~\cite{Park2019SpecAugmentAS}, using 83-dim log-mel filterbank features with pitch. The encoder consists of 2 convolutional layers and 12 self-attention blocks, and the decoder contains 6 self-attention blocks. The training loss jointly minimizes sequence-to-sequence (S2S) and connectionist temporal classification (CTC) objectives~\cite{Kim2017JointCB}. The decoder predicts subword units generated using SentencePiece~\cite{Kudo2018SentencePieceAS}. For decoding, we used beam search which combines scores from the S2S, CTC, and an external Transformer-based language model. This ASR model obtains WERs of 2.2\% and 5.5\% on the Librispeech \texttt{test-clean} and \texttt{test-other} evaluation sets, respectively, using a beam size of 60.

\subsection{Experimental results}
\vspace{-0.2em}

Similar to our diarization experiments, we evaluated the ASR models on mixed and separated audio. Table~\ref{tab:mixed_asr} shows the results obtained on the mixed LibriCSS data. For these experiments, we used oracle segments and speaker information. From the table, we see that a strong Librispeech model also performed well on mixed LibriCSS utterances. Among the TDNNF-based hybrid models, fine tuning on reverberated data provided a 31.5\% relative WER improvement. For transformer-based E2E models, decoding with larger beams (of size 30) improved WER by almost 10\% relative, compared to decoding with smaller beam sizes. We did not get any significant gains by increasing beam sizes further, so we used this setting for further experiments.

Next, we present ASR results in the context of our pipeline, i.e., using separated audio streams from the mask-based MVDR model and segments from spectral clustering based diarization. We used the \textit{fine-tuned} variant of the hybrid HMM-DNN ASR model. Additionally, to emphasize the importance of the separation module, we also show cpWERs obtained on mixed recordings. The results are shown in Table~\ref{tab:asr_sep}. On replacing oracle segments with those obtained from diarization, the downstream cpWERs for hybrid and E2E ASR systems degraded by 23.4\% and 37.6\%, respectively. Interestingly, this increase was more prominent in the low and medium overlap conditions, which suggests that errors in these cases occured primarily from incorrect speaker assignment. On applying separation, the cpWERs improved significantly (although this was partly due to better diarization). In particular, we found that the E2E model benefited more from separated audio streams, with its cpWER improving from 27.1\% to 13.4\%. For both the models, the cpWER with separation was found to outperform the corresponding results on mixed recording even using oracle segments. 

Finally, with all our experimental evaluations in place, we combined the best performing models at each stage of the pipeline. We prepared two variants: (a) without separation, and (b) with separation. For (a), we selected TS-VAD based diarization and the Transformer-based ASR model. Pipeline (b) consists of sequential multi-frame separation, followed by SC-based diarization and the same ASR. We present the final results for both variants in Table~\ref{tab:final}. It is evident that in the absence of an explicit separation module, even a well performing diarization and ASR combo is hamstrung. We discuss some more implications of these results in the next section.

\begin{table}[t]
\centering
\caption{Pipeline variants: (a) without separation, and (b) with separation, combining best models at each stage.}
\label{tab:final}
\begin{adjustbox}{max width=\linewidth}
\begin{tabular}{@{}llllcc@{}}
\toprule
 & \textbf{Separation} & \textbf{Diarization} & \textbf{ASR} & \textbf{DER(\%)} & \textbf{cpWER(\%)} \\ \midrule
\textbf{(a)} & - & TS-VAD & Transformer & \textbf{7.6} & 23.9 \\
\textbf{(b)} & Sequential & SC & Transformer & 14.1 & \textbf{12.7} \\ \bottomrule
\end{tabular}
\end{adjustbox}
\end{table}

\vspace{-1em}
\section{Discussion}
\vspace{-0.2em}

Our extensive experiments using a variety of separation, diarization, and ASR methods elicit important lessons for solving the multi-speaker speech recognition problem. First, the importance of explicit separation cannot be overstated --- the best cpWER for a pipeline with separation is 46.9\% relative better than one without it (Table~\ref{tab:final}). For separation, we found training and inference with larger chunks to perform better; there is a caveat, however --- the improved performance is obtained through diarization tricks that can filter out the increased cross-talk with these models. Second, we observed that although new (supervised) diarization methods like RPN and TS-VAD provide substantial gains on mixed recordings, the performance does not carry over well to separated audio streams, and traditional clustering approaches can still outperform them in these settings (Table~\ref{tab:sep_diar}). In particular, the best diarization  without separation was found to be 45.8\% relative better than one with separation, which is highly counter-intuitive. This is particularly relevant for pipelines such as ours and the CSS pipeline, where separation is performed before diarization, and we regard this as an important direction for further investigation. There have also been concurrent efforts to apply diarization on separated audio streams~\cite{Xiao2020MicrosoftSD}. Finally, unlike diarization, ASR performance on mixed recordings was strongly indicative of the performance on separated audio streams. We found that a state-of-the-art ASR model trained on clean, single speaker utterances integrates well in the pipeline and results in the best available cpWER on this dataset. This is encouraging, especially in view of recent advances in end-to-end speech recognition.

\small
\textbf{Acknowledgment.} The work reported here was started at JSALT 2020 at JHU, with support from Microsoft, Amazon, and Google.


\bibliographystyle{IEEEbib}
\bibliography{strings,refs}

\end{document}